\documentclass[12pt]{article}
\newcommand{\la}[1]{\label{#1}}

\newcommand{\vecn}{{{\bf n}}}

\newcommand{\vect}{{{\bf t}}}

\newcommand{\vecb}{{{\bf b}}}
\newcommand{\vecx}{{{\bf X}}}
\newcommand{\vece}{{{\bf e}}}

\newcommand{\be}{\begin{equation}}
\newcommand{\ee}{\end{equation}}
\newcommand{\ba}{\begin{eqnarray}}
\newcommand{\ea}{\end{eqnarray}}
\newcommand{\bastar}{\begin{eqnarray*}}
\newcommand{\eastar}{\end{eqnarray*}}
\newcommand{\half}{{1 \over 2}}
\begin{document}
\begin{titlepage}

\vskip 2.0truecm

\begin{center}
$ ~$
\end{center}

%\vskip 1.0cm

\begin{center}
{
\bf \large \bf  PHASES OF BOSONIC STRINGS AND
\\ \vskip 0.2cm
TWO DIMENSIONAL GAUGE THEORIES  \\
}
\end{center}

\vskip 0.5cm

\begin{center}
{\bf Antti J. Niemi} \\
\vskip 0.6cm

{\it Department of Theoretical Physics,
Uppsala University \\
P.O. Box 803, S-75108, Uppsala, Sweden } \\

\end{center}

\vskip 2.0cm
\noindent
We suggest that the extrinsic curvature and torsion
of a bosonic string can be employed as
variables in a two dimensional Landau-Ginzburg
gauge field theory. Their interpretation 
in terms of the abelian Higgs multiplet leads to
two different phases. In the phase with unbroken 
gauge symmetry the ground state describes 
open strings while in the phase with broken gauge symmetry 
the ground state involves closed strings. When 
we allow for an additional abelian gauge structure along 
the string, we arrive at an interpretation in terms
of the two dimensional SU(2) Yang-Mills theory.

\vskip 2.0cm
\vfill

\begin{flushleft}
\rule{5.1 in}{.007 in} \\
{\small  E-mail: \scriptsize
\bf NIEMI@TEORFYS.UU.SE  } \\
\end{flushleft}

\end{titlepage}

In the Wilsonian approach to renormalization 
group equations the Polyakov action \cite{poly1} 
is a relevant term for describing the high energy 
limit of a bosonic string. But at lower energies there 
can be corrections such as the extrinsic curvature
term \cite{poly2}. Here we 
shall be interested in additional corrections
that emerge from the extrinsic geometry of the string 
and that may affect its low energy behaviour. 
For definiteness we consider a bosonic string 
in three spatial dimensions; Even though the 
quantization of the Polyakov action dictates D=26, 
the role of a critical dimension becomes less 
obvious when higher order corrections are included.
This is because such corrections in general fail 
to be conformally invariant on the worldsheet. 
We shall suggest that in three 
spatial dimensions the extrinsic geometry leads to 
effective two dimensional Landau-Ginzburg
gauge field theories, such as the abelian 
Higgs model and the SU(2) Yang-Mills theory. The abelian 
Higgs model with its two distinct phases is particularly 
interesting. We propose that the phase with unbroken gauge
symmetry is natural for describing open strings. 
The phase where the U(1) gauge symmetry becomes 
spontaneously broken is then more natural for describing
closed strings. In this phase the Polyakov action 
admits an interpretation
as the vacuum expectation value of the Higgs field, it gives
a mass to the two dimensional vector field. 
Finally,  we consider the extrinsic geometry
of a bosonic string when equipped with an additional U(1) 
gauge structure. Now the quantities that describe 
the string can be combined into the two dimensional SU(2)
Yang-Mills field.

We start from the static limit of the 3+1 dimensional 
classical bosonic string action
\be
S \ = \ \mu^2 \hskip -0.1cm \int d\sigma dt \sqrt{-g} g^{ab} 
\partial_a X^\mu \partial_b X^\mu \ \ \ \ \ \ (\mu = 0,1,2,3)
\la{poly1}
\ee
We choose a proper time gauge where the world sheet
coordinate $t$ becomes proportional to
$X^0$, thus in the static limit the $t \propto X^0$ dependence 
can be ignored. The remaining spatial variables
$X^i(\sigma)$ ($i=1,2,3$) then 
describe the string as a curve which is embedded in $R^3$. 
We choose $\sigma$ so that it coincides with the
$R^3$ arc-length of the string, $\sigma \to s \in [0,L]$. 
Here $L$ is the (variable) total 
length of the string in $R^3$, a remnant of the 
modular parameters on the string world sheet. 
The static energy now reduces to 
\be
E \ = \ \mu^2 \int\limits_0^L ds \partial_s X^i 
\partial_s X^i \ = \ \mu^2 \cdot L
\la{poly2}
\ee
Note that the energy of the string depends only on 
the string tension $\mu^2$ and the
(modular) length $L$.  

The three component unit vector
\be
\vect \ = \ \frac{d \vecx}{ds}
\la{fret}
\ee
is tangent to the curve $X^i(s)$ in $R^3$. Together with 
the unit normal $\vecn$ and the unit
binormal $\vecb = \vect \times \vecn$ we then  
have an orthonormal frame in $R^3$, at each point 
along the string. In terms of the complex
combination $\vece_F^\pm = \half ( \vecn \pm i
\vecb)$ these vectors are subject to the Frenet 
equations 
\ba
\frac{d\vect}{ds} \ & = & \ \half \kappa (\vece_F^+ + \vece_F^-) \\
\frac{d\vece_F^\pm}{ds} \ & = & \ - \kappa \vect \mp i \tau 
\vece_F^\pm
\la{frenet}
\ea
Here $\kappa$ is the extrinsic curvature and
$\tau$ is the torsion of the string,
\be
\kappa \ \equiv \ \kappa_\pm \ = \ \vece_F^\pm 
\cdot \partial_s \vect
\la{kappa}
\ee
\be
\tau \ = \ \frac{i}{2} \vece_F^- \cdot \partial_s 
\vece_F^+
\la{tau}
\ee
Notice that we write $\kappa_\pm$ to distinguish the
two different but equivalent representations of the 
curvature in terms of the complex 
Frenet frame ($\vect, \vece_F^\pm)$.
Obviously any physical property of the string should be
independent of the choice of basis vectors on the normal planes.
Instead of the Frenet combination $\vece_F^\pm$ we
may then introduce an arbitrary complex orthonormal frame 
$\vece^\pm$ that relates to the Frenet frame $\vece_F^\pm$ by 
a rotation with an angle $\theta(s)$ on the 
normal planes,  
\be
\vece_F^\pm \ \to \ \vece^\pm \ = \ e^{\pm i \theta} \vece_F^\pm
\la{rota1}
\ee
This rotation sends $\kappa_\pm$ to the complex conjugate
pair  
\be
\kappa_\pm \ \to \ e^{\pm i \theta} \kappa_\pm
\la{rotakappa}
\ee
while for the torsion we get
\be
\tau \ \to \tau \ - \ \partial_s \theta
\la{rotatau}
\ee
In (\ref{rotakappa},\ref{rotatau}) we identify
the gauge transformation 
structure of a two dimensional abelian Higgs multiplet 
($\phi, A_i$): The frame rotation (\ref{rota1}) 
corresponds to a static U(1) gauge transformation, 
$\kappa_+$ together with its complex conjugate $\kappa_-$
corresponds to the complex scalar field 
$\phi \sim \kappa_+$ and $\tau$ to the spatial 
component $A_1 \sim \tau$ of the U(1) gauge field
in the static $A_0 = 0$ gauge. 
Since the physical 
properties of the string are independent of 
the local frame, they should remain invariant 
under the U(1) transformations (\ref{rotakappa},\ref{rotatau}). 
In partcular any Landau-Ginzburg
energy of the string which involves the 
multiplet ($\kappa_+, \tau$) $\sim$ ($\phi, A_1$) 
should be U(1) gauge invariant. 

The abelian Higgs model admits two different phases.
Consequently we have two alternatives 
for the static U(1) invariant 
Landau-Ginzburg energy. We first consider the phase 
where the gauge symmetry remains unbroken. It is described 
by the following Landau expansion
\be
E_I \ = \ \mu^2 \int\limits_0^L ds (\partial_s X^i)^2
\ + \ \int\limits_0^L ds \left\{ \alpha |\phi|^2 \ +
\ \beta |(\partial_s + i A_1 ) \phi|^2 \ + \ ... \ \right\}
\la{phase1a}
\ee
In the ground state the average value of local curvature
along the string vanishes,
\be
<|\phi|^2> \ = \ 0
\la{min1}
\ee
Since $L\not= 0$ this means that in the ground state
the string can not form a closed curve. Consequently 
this phase is appropriate
for describing open strings. Note that the first two 
terms in (\ref{phase1a}) reproduce the rigid string action of 
\cite{poly2}, and the third term 
is a higher derivative correction. Consequently we
expect that (\ref{phase1a}) describes 
strings in the same universality class with the 
action in \cite{poly2}.

The phase where the gauge symmetry becomes broken by the
Higgs effect can be described by the following Landau-Ginzburg
energy,
\be
E_{II} \ = \ \int\limits_0^L ds \left[ |(\partial_s + i A_1 ) 
\phi|^2 \ + \ \lambda ( |\phi|^2 - a^2 )^2 \right]
\la{phase2a}
\ee
Notice that unlike in (\ref{phase1a}), now we do  
not introduce (\ref{poly1}) explicitely. This term 
is already contained in the field independent part of the 
potential in (\ref{phase2a}) which according to
(\ref{poly2}) can be rewritten as
\[
\int\limits_0^L ds \ \lambda ( |\phi|^2 - a^2 )^2
\ = \ \int\limits_0^L ds \ \lambda \left[ |\phi|^4 - 
2 a^2 |\phi|^2 \sqrt{ \partial_s X^i \partial_s X^i } 
+ a^4 \partial_s X^i \partial_s X^i \right]
\]
In particular the (static) string action (\ref{poly1}) 
is proportional to the ground state expectation value 
of the complex Higgs scalar $\phi$,
\be
<|\phi|^2> \ = \ a^2  \sqrt{ \partial_s X^i \partial_s X^i }
\la{min2}
\ee
With $a^2 \not= 0$ we have spontaneous symmetry breaking, 
and (\ref{phase2a}) 
describes the string in a phase which is different 
from that described by (\ref{phase1a}). Notice that in the
ground state (\ref{min2}) the energy (\ref{phase2a}) vanishes. 
In particular, this ground state energy is independent 
of the length (modular parameter) $L$.
The U(1) gauge invariant variables $(\rho, C$) 
are defined by
\be
\phi = \rho e^{i \chi} \ \ \ \ \ \ \& \ \ \ \ \ \ \tau = C + 
\partial_s \chi
\la{givar}
\ee
In terms of these we have for the energy
\be
E_{II} \ = \ \int\limits_0^L ds \left\{ (\partial_s 
\rho)^2 + \rho^2 C^2 + \lambda (\rho^2 - a^2 ) \right\}
\la{giact}
\ee
The variable $\rho$ describes the curvature of the 
string, and $C$ describes the (frame independent) torsion. 
There is an interplay between the parameters $L$ and $a$ and the
ground state geometry of the string: When $L\cdot a 
= 2\pi n $ the ground state of the string has 
$\rho^2 = a^2$ and $C=0$ which corresponds to a circular
planar unknot in $R^3$. For other values of $L\cdot a$ 
the ground state geometry becomes more involved. With 
$L \cdot a \geq 2\pi$ the ground state string can form
a closed curve in $R^3$, and the Landau-Ginzburg 
energy (\ref{phase2a}) is appropriate 
for describing closed strings. But for $0 < 
L\cdot a < 2\pi$ the ground state string with vanishing 
energy is an open, bent string.
 
The phase factor $\chi$ in (\ref{givar}) is defined modulo 
$2\pi k$, with $k$ an integer. In the abelian Higgs model
this integer labels the different instanton vacua.
In the present case $k$ counts the number
of times the (normal frame of the) string rotates around its
axis when we move once around the string. Since $\chi$ 
is absent in (\ref{giact}) these rotations of the normal frame
have no effect to the energy. In order to relate the strings with a 
different $k$ with the instanton vacua, we add
to the arc-length $s$ an additional coordinate $t$ which we
view as the Euclidean time coordinate and extend (\ref{phase2a})
to the abelian Higgs model 
\be
S \ = \ \int d^2x \left[ \ \frac{1}{4} F_{\mu\nu}^2
+ |(\partial_\mu + i A_\mu) \phi |^2 + \lambda (|\phi|^2 
- a^2 )^2 \ \right]
\la{relact1}
\ee
We interpret this as the (covariantized) Euclidean time dependent
action for the string worldsheet, it clearly 
reduces to (\ref{phase2a}) for static strings in the 
$A_0 = 0$ gauge. This action supports two dimensional
vortices as instantons. We assume
that $L \cdot a = 2\pi$ so that the ground state string is a 
circular planar unknot; it corresponds to a vacuum state
of the Higgs action (\ref{relact1}). We introduce an instanton that 
interpolates between two such ground 
state string vacua at $t = \pm T \to \pm \infty$ with
different integers $k$.  
The instanton has a 
nontrivial first Chern character (= integer $m$), 
and in the $A_0 = 0 $ gauge
\be
Ch_1 (F) \ = \ \frac{1}{2\pi} \int F \ = \ \frac{1}{2\pi}
\oint_{+T} ds \tau (s) \ - \ \frac{1}{2\pi}
\oint_{-T} ds \tau(s) \ = \ m
\la{1st}
\ee
This coincides with the difference in the integers
$k$ that count the number of times the frames of the
ground state strings at $t=\pm T$ rotate around their axis
when we move once around the strings.

In general the string can self-link, and in particular 
a closed string can form a knot. The 
Calugareanu theorem \cite{calu} 
states that the (integer valued) self-linking 
number $Lk$ of a knotted string $K$ equals the sum of its 
twist $Tw$ and writhe $Wr$,
\[
Lk(K) \ = \ Tw(K) \ + \ Wr(K)
\]
This resembles an index theorem for the instanton
in the abelian Higgs model (\ref{relact1}). For this
we consider a Seifert surface ${\cal S}$ of $K$. This 
a smooth orientable Riemann surface in $R^3$ 
with one boundary component that coincides with the 
string $K$. We re-interpret (\ref{relact1}) as a 
(static) Hamiltonian on the Seifert surface, 
with $\kappa_\pm$ and $\tau$ 
appropriately extended into an abelian Higgs multiplet 
$(\phi, A_i)$ on ${\cal S}$ so that $\tau$ 
coincides with the component of $A_i$ which is tangential 
to $K$ along the boundary of ${\cal S}$, and $\phi$ is equal
to $\kappa_+$ on the boundary. We define a two 
dimensional Dirac operator $\slash \hskip -0.28cm 
D = \gamma^a e_a^i(\partial_i + A_i + \omega_i)$ 
on ${\cal S}$, with $\omega_i$ the 
spin connection. The Atiyah-Patodi-Singer index theorem
computes the index of this Dirac operator on 
${\cal S}$ in terms of the first 
Chern character (\ref{1st}) over ${\cal S}$ and 
the $\eta$ invariant of the restriction of 
$\slash \hskip -0.28cm D $ on the
boundary of ${\cal S}$,
\[
{ \rm Index } \slash {\hskip -0.28cm D} \ = \ 
\frac{1}{2\pi} \int_{\cal S} F \ + \ \half
\eta \ = \ \frac{1}{2\pi} \oint_K \tau \ + \
\half \eta
\]
Presumably this coincides with the Calugareanu relation:
The self-linking of the string
equals the index of the Dirac operator, its twist 
equals the integral of $F = dA$ over the
Seifert surface, and the writhe coincides with the 
$\eta$-invariant of the boundary Dirac operator. 
Notice that the twist can also be interpreted physically,
as the magnetic flux through the Seifert surface.

Similarly various other properties of the abelian Higgs
model can be interpreted in terms of string geometry.
Instead we now proceed to a relation
with the two dimensional SU(2) Yang-Mills field.
For this we again consider the evolution
of a string in $R^3$, amending the arc-length $s$
with a Euclidean (time) coordinate $t$. We extend 
$\kappa_\pm$ and $\tau$ into this two dimensional 
space by interpreting {\it both} as (spatial) 
$s$-components of two dimensional vector fields,
$\kappa^\pm_i$ and $\tau_i$ ($i=s,t)$ respectively. The 
$t$-components of these vector fields are
obtained by considering the $t$-evolution of the
local frame vectors ($\vect, \vecn, \vecb$)
and setting    
\be
\kappa_\pm \ = \ \vece_\pm \cdot \partial_s \vect \ \to
\ \kappa^\pm_i \ = \ \vece_\pm \cdot \partial_i \vect
\la{kappa2}
\ee
\be
\tau \ = \ \frac{i}{2} \vece_- \cdot \partial_s \vece_+
\ \to \ \tau_i \ = \ \frac{i}{2} \vece_- \cdot \partial_i
\vece_+
\la{tau2}
\ee
The $t$-evolution of the string parametrizes a
two dimensional worldsheet $X^i(s,t)$ in $R^3$,
determined by the motion of 
an initial string $K_1$ at $t_1$ into a 
final string $K_2$ at $t_2$. For each $t\in (t_1,t_2)$ 
the tangent vector $\vect(s,t)$ maps the ensuing curve
$X^i(s,t)$ onto a unit circle $S^1 \in R^3$. The 
$t$ dependence of $\vect(s,t)$ then parametrizes a family of 
such circles on the surface of a 
two-sphere $S^2_\vect \in R^3$, 
for which $\vect(s,t)$ is a unit normal. With $u^a$ ($a=1,2$) 
local coordinates on $S^2_\vect$, the one-form
\be
d \vect \ = \ - \partial_a \vecx {B^a}_{b} du^b  
\la{defcurv}
\ee
defines the components of the (symmetric) extrinsic 
curvature tensor ${B^a}_b$ for the surface which is
drawn by $\vect(s,t)$ on the sphere $S^2_\vect$. Since
$S^2_\vect$ is a unit two-sphere, the 
two eigenvalues {\it i.e.} principal curvatures $\beta_a$ 
of the extrinsic curvature tensor ${B^a}_b$ 
coincide, that is $\beta_1 = \beta_2 = 1$. 

We now generalize the present construction, by interpreting
$\vect$ as a unit normal of an arbitrary (genus 0) two dimensional
surface ${\cal G} \in R^3$ instead of the sphere $S^2_\vect$. 
The curvature tensor ${B^a}_b$ of ${\cal G}$ is an
arbitrary symmetric matrix so that its eigenvalues, 
the two principal curvatures $\beta_a \to \lambda_a$ of ${\cal G}$
are also arbitrary,
\[
diag({B^a}_b) \ = \ \left( \begin{array}{cc} 1 & 0 \\ 0 & 1 
\end{array}\right) \ \rightarrow \  \left( \begin{array}{cc} 
\lambda_1 & 0 \\ 0 & \lambda_2 
\end{array}\right) \
\] 
We combine these eigenvalues of the extrinsic curvature 
tensor on
${\cal G}$ into the complex $\lambda_\pm = \lambda_1 
\pm i \lambda_2$ and generalize (\ref{kappa2}) from $S^2_\vect$
into ${\cal G}$ which amounts to
\be
\kappa^\pm_i \ = \vece^\pm \cdot \partial_s \vect 
\ = \ \vece^\pm \cdot \partial_a \vecx {B^a}_b \partial_s u^b
\ \to \ \psi^\pm_i \ = \ \lambda_\pm 
\kappa^\pm_i
\la{kappa2b}
\ee
We now have an additional U(1) stucture, 
corresponding to frame rotations on the tangent 
plane of ${\cal G}$. This sends 
\be
\lambda_\pm \ \to \ e^{\pm i \varphi} \lambda_\pm
\la{rota2}
\ee
The vector $\psi^\pm_i$ is then invariant under a
diagonal U$_+$(1) $\in$ U(1)$\times$U(1) transformation 
where we compensate a frame rotation (\ref{rota1})
with an opposite frame rotation (\ref{rota2}) on 
${\cal G}$, with $\theta + \varphi = 0$. 
But $\psi^\pm_i$ is a charged vector 
field under the remaining U$_-$(1) $\in$ 
U(1)$\times$U(1), with angle $\theta - \varphi$.
A Landau-Ginzburg action of the string should be 
invariant under these frame rotations. For 
this we introduce a gauge vector $C_i$ such that
$C_i \to C_i - \partial_i \varphi$
under (\ref{rota2}).  We then define   
\be
B_i \ = \ C_i + \tau_i
\la{bi}
\ee
The ($\psi^\pm_i, B_i$) is then a charged
vector multiplet which is invariant under the U$_+$(1) 
combination of frame rotations and gauged under the 
U$_-$(1) combination of frame rotations.
Notice the dual structure between the two U(1) 
frame rotations, in particular the 
multiplets ($\tau_i, \kappa^\pm_i$) and ($C_i , \eta^\pm$) 
can be viewed as electric and magnetic dual variables. 

We denote
\be
A^3_i \ = \ B_i \ \ \ \ \ \ 
\& \ \ \ \ \ \ A^\pm_i \ = \ \psi^\pm_i
\la{ym1}
\ee
where the ($3,\pm$) refer to the SU(2) Lie algebra
in the Cartan basis. This defines a two-dimensional SU(2) 
Yang-Mills multiplet which involves six independent field degrees 
of freedom. This equals the number of field
degrees of freedom in a generic 
two dimensional SU(2) Yang-Mills gauge field, 
consequently we identify (\ref{ym1}) as the (decomposed) 
components of the full D=2 SU(2) Yang-Mills field; see \cite{fadde}.
In particular, the two dimensional Yang-Mills action can now
be viewed as a Landau-Ginzburg action to describe the extrinsic 
geometry of a 3+1 dimensional (closed) bosonic string that
admits an additional U(1) structure which emerges when
we extend $S^2_\vect$ to ${\cal G}$.

In conclusion, we have investigated the extrinsic geometry
of bosonic strings in 3+1 dimensions. In particular, we have 
proposed that the extrinsic curvature and torsion can be viewed as
variables in a two dimensional gauge field theory. This yields 
the abelian Higgs model as a Landau-Ginzburg description
of the string, with its two phases relating to open and closed 
strings in a rather natural fashion. Furthermore, if we allow 
for an additional U(1) structure along the string, we can also 
interpret the two dimensional SU(2) Yang-Mills theory as a 
Landau-Ginzburg action of a 3+1 dimensional strings.
  
\vfill\eject

We thank T. Jonsson, U. Lindstr\"om, L. Thorlacius and K. Zarembo
for discussions. We also thank University of Iceland for hospitality 
during the completion of this article. This research has been
partially supported by The Swedish Research Council grant number 
F-AA/FU 06821-308

\end{document}